\documentclass[%
 aip,
 pop,%
 amsmath,
amssymb,
reprint,%
]{revtex4-1}
\usepackage{graphicx}%
\usepackage{dcolumn}%
\usepackage{bm}%
\usepackage[latin1]{inputenc}
\usepackage{graphicx,color}
\begin{document}

%\preprint{}

\title{Turbulence driven particle transport in Texas Helimak} %Title of paper



\author{D. L. Toufen}
%\thanks{}
\altaffiliation{dennis@if.usp.br}
\affiliation{Institute of Physics, University of São Paulo, C.P. 66318, 05315-970 São Paulo,
São Paulo, Brazil}
\affiliation{Federal Institute of Education, Science and Technology of São Paulo - IFSP, 07115-000 Guarulhos,
São Paulo, Brazil}

\author{Z. O. Guimarães-Filho}
%\thanks{}
\affiliation{Aix-Marseille Univ., CNRS PIIM UMR6633, International Institute for Fusion Science, Marseille, France}

\author{I. L. Caldas}
%\thanks{}
\affiliation{Institute of Physics, University of São Paulo, C.P. 66318, 05315-970 São Paulo,
São Paulo, Brazil}

\author{F. A. Marcus}
%\thanks{}
\affiliation{Aix-Marseille Univ., CNRS PIIM UMR6633, International Institute for Fusion Science, Marseille, France}

\author{K. W. Gentle}
%\thanks{}
\affiliation{Department of Physics and Institute for Fusion Studies, The University of Texas at Austin,
Austin, Texas 78712, USA}


\date{\today}

\begin{abstract}
We analyze the turbulence driven particle transport in Texas Helimak (K. W. Gentle and Huang He, Plasma Sci. and Technology, 10, 284  (2008)), a toroidal plasma device with one-dimensional equilibrium with magnetic curvature and shear. Alterations on the radial electric field, through an external voltage bias, change spectral plasma characteristics inducing a dominant frequency for negative bias values and a broad band frequency spectrum for positive bias values. For negative biased plasma discharges, the transport is high where the waves propagate with phase velocities near the plasma flow velocity, an indication that the transport is strongly affected by a wave particle resonant interaction. On the other hand, for positive bias the plasma has a reversed shear flow and we observe that the transport is almost zero in the shearless radial region, an evidence of a transport barrier in this region.
\end{abstract}


%Use showkeys class option if keyword

\maketitle %\maketitle must follow title, authors, abstract and \pacs

\section{Introduction}
%\label{}

Improvement in magnetically confined plasmas in toroidal devices, as tokamaks and stellarators, is limited by the anomalous particle transport at the plasma edge [\onlinecite{Hidalgo04}] driven by the electrostatic turbulence observed at this region  [\onlinecite{Horton99},\onlinecite{Garbet01}]. In the last years, several works have investigated the turbulence and transport reduction by imposing an external electric potential that changes the radial electric field profile [\onlinecite{Hidalgo05,Nascimento05,VanOost03}]. 
However, more observations are required to complete the experimental description and identify the proper theoretical concepts that could lead to a complete interpretation of the transport control achieved by modifying the electric field. Thus, several experiments have been performed in tokamaks and stellarators.  

Observations in other type of magnetic confinements can also help to expand the experimental description of the effects of the radial electric profile on the turbulence and the transport. Thus, to improve the experimental knowledge of basic turbulence and transport, experiments have been performed in some devices which confine plasmas with selected characteristics of a fusion plasma in a simpler geometry and with better diagnostics than those possible in major confinement devices [\onlinecite{Rypdal05,Perez06,Ricci08}].

These basic plasma devices can present a cylindrical geometry as the LAPD  [\onlinecite{Gekelman91}] which was used to study the influence of the bias in the turbulence and transport [\onlinecite{Carter09}]. Other machines with toroidal geometry present the magnetic field line curvature in combination with plasma gradients as present in fusion experiments. One possibility for these plasma toroidal devices is the helimak configuration, present in the BLAAMANN [\onlinecite{Rypdal94}], TORPEX [\onlinecite{Fasoli03}], and Texas Helimak [\onlinecite{Gentle08}], which has been used to study electrostatic instabilities, turbulence, and transport[\onlinecite{Muller05},\onlinecite{Fasoli06}]. The helimak configuration is a basic plasma toroidal device with characteristics of a fusion plasma in a simple geometry [\onlinecite{Gentle08},\onlinecite{Dahlburg09},\onlinecite{Luckhardt99}], with a sheared cylindrical slab that simplifies the turbulence description and provides results that can be used to understand plasma edge and the scrappe-of layer transport in major fusion machines. Altough a stationary equilibrium is not expected [\onlinecite{Tasso03}] in this configuration, a turbulent state with a peaked radial pressure profile exists [\onlinecite{Rypdal94}]. As the plasma in helimaks is colder and less dense when compared with tokamaks, it is possible to use a large set of diagnostic probes.

The Texas Helimak is also capable of producing states of greatly reduced turbulence by biasing and has an independent spectrocopic diagnostic of the plasma flow. These features make the Texas Helimak an interesting device to study the plasma flow shear influence on the particle transport. Some of the analyzed discharges present a reversed shear plasma flow and are especially adequate to investigate transport barrier onset predicted to this kind of flow [\onlinecite{Negrete00,Sommeria89,Solomon93}].

In this article we analyze the turbulence driven particle transport in Texas Helimak and investigate how alterations on the radial electric field, through an external voltage bias, modify the turbulence and the transport. We employ spectral analysis to identify the main changes on the power spectra due to the external alterations on the radial electric field profile. Moreover, the biasing also allows observing the dependence of turbulence spectrum and transport on the flow velocity shear. 

Thus, we calculate the transport radial profile and its dependence on the electric bias. We observe that the external bias value changes spectral plasma characteristics, inducing a dominant frequency for  negative bias values and a broad band frequency spectrum for positive bias values. Because of this difference the shots with negative and positive bias values are analyzed separately. For negative biased shots, the transport reaches a maximum where the waves propagate with phase velocities approximately equal to the plasma flow velocity. We interpret this maximum as evidence that the transport is strongly affected by a wave particle resonant interaction. On the other hand, for positive bias the plasma has a reversed shear flow and we observe that the transport is almost zero in the shearless radial region indicating an evidence of a transport barrier in this region.

\section{Experimental set up}
%\label{}

\begin{figure}
	\centering
				\includegraphics[]{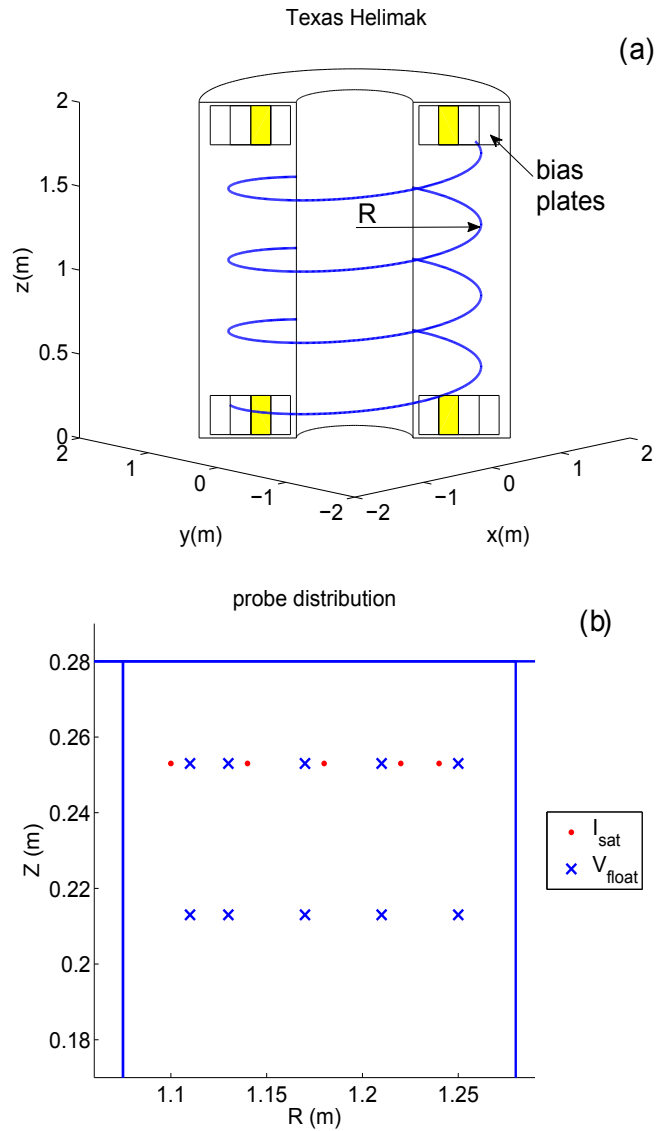}
	\caption{(a) Cross section of the Texas Helimak vacuum vessel showing the position of the four sets of bias plates (in yellow are those used for this work) and a sample of its helicoidally magnetic field lines.  (b) One of the probe distributions used in this work: the red dots represent the probes used to measure the saturation current, the blue cruxes the probes used to the floating potential and the blue lines mark the bias plate position.}
	\label{fig:helimak}
\end{figure}

We analyze experiments performed at the Texas Helimak [\onlinecite{Gentle08}], Fig. \ref{fig:helimak} (a), a basic plasma toroidal device located at the University of Texas at Austin. In this machine, the combination between the toroidal and the small vertical field creates a helicoidally geometry of its magnetic field lines with curvature and shear. The Helimak geometry is an approach to a sheared cylindrical slab [\onlinecite{Luckhardt99}] since connection lengths are long enough to neglect the end effects. Most of these magnetic field lines starts and terminate into four sets of four plates located at 180° apart at the top and the bottom part of the machine. These plates are used as a support to the Langmuir probes and to apply external electric potentials (bias) to changes the radial electric field profile. In the analyzed experiments, the dominant toroidal field is about 0.1 T, which addicted with the weaker vertical field create magnetic field lines with $\approx$ 40m of connection length at the middle of the machine (R = 1m). The field lines are thus helices as shown in Fig. \ref{fig:helimak} (a), spiraling from bottom to top.

Texas Helimak has a vacuum vessel with rectangular cross section with external radius, R$_{external}$ = 1.6 m, internal radius, R$_{internal}$ = 0.6 m, and height = 2 m. For the experiments analyzed in this work, Argon gas at $10^{-5}$ Torr was heated by ECRH with 6kW of power inserted by a window located on the inner side of the vacuum vessel. The shot duration is up to 20 s and the plasma is in a steady state with stationary conditions during 10 s, the time interval considered for fluctuation analyzes described in this work.

The diagnostic system count with more than 700 Langmuir probes mounted at the four sets of bias plates. The analyzed data were taken by two digitizers, one with 16 channel and 100 kHz of sample rate and another one with 64 channels and 7 kHz of sample rate. One of the used probe distributions is shown at Fig. \ref{fig:helimak} (b), where the red dots represent the probes used to measure the ion saturation current, blue cruxes the probes used to the floating potential, and the blue lines indicate the bias plate position. This configuration is used to measure the turbulence induced radial transport. 

\begin{figure}
	\centering
				\includegraphics{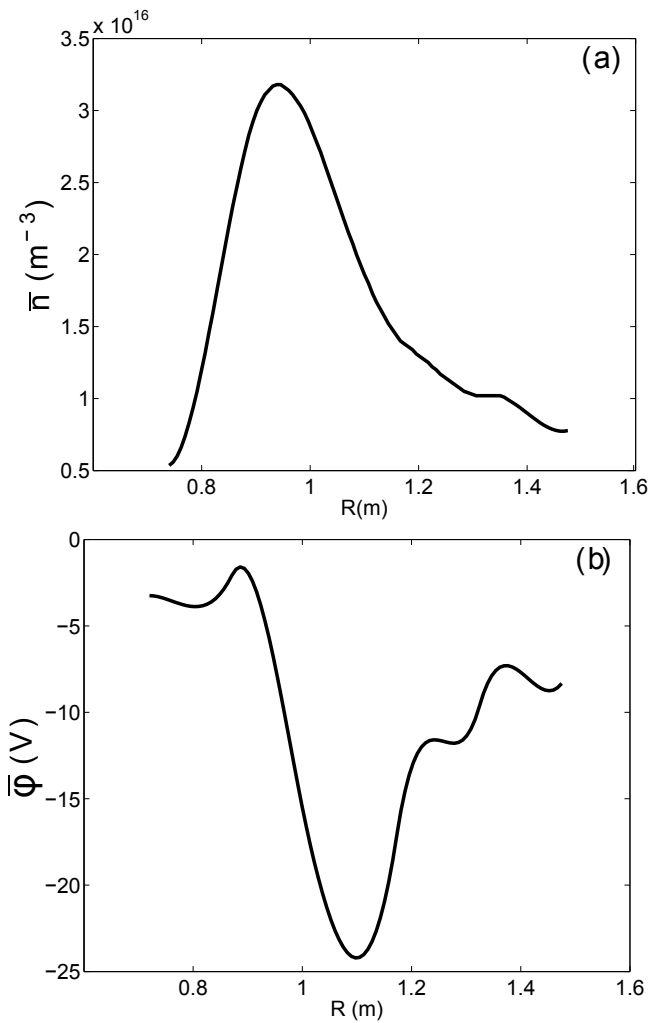}
	\caption{Radial profiles of (a) mean plasma density estimated from current saturation and (b) mean floating potential.}
	\label{fig:perfis_medios}
\end{figure}

Figure \ref{fig:perfis_medios} (a) shows the mean density radial profile estimated from saturation current fluctuations. The density profile has a maximum at R $\approx$ 0.95 m and decreases more smoothly on the external side of the peak, which corresponds to the low field side of the toroidal magnetic field.  Figure \ref{fig:perfis_medios} (b) shows the mean floating potential radial profile which presents a minimum at R $\approx$ 1.1 m and a positive gradient in the analyzed radial region, 1.1m $<$ R $<$ 1.25m. In this low field side region, the density gradient is practically uniform.

\begin{figure}
	\centering
				\includegraphics{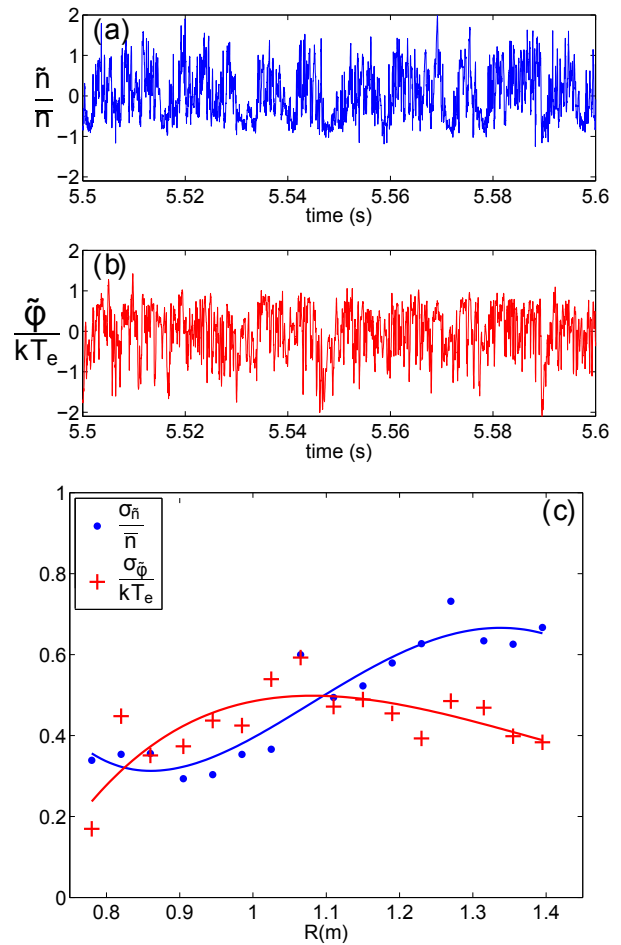}
	\caption{Evolution of normalized fluctuation of (a) ion saturation current and (b) floating potential in a shot without biasing. The saturation current signal was measured at R = 1.14m and the floating potential at R = 1.13m.  (c) Radial profile of the density (blue dots) and floating potential (red crosses) turbulent level. The lines are guides to the eyes.}
	\label{fig:flutuacoes}
\end{figure}

We show in Fig. \ref{fig:flutuacoes} (a) and (b) examples of evolution of the normalized saturation current and floating potential during a short interval of 0.1s (of the 10 s considered for fluctuation analyzes) in a shot without biasing. The saturation current signal presented in Fig. \ref{fig:flutuacoes}(a) was measured at R = 1.14m and the floating potential of Fig. \ref{fig:flutuacoes}(b) at R = 1.13m (both in the radial interval chosen for fluctuation analyses). For the normalization of the floating potential presented in Fig. \ref{fig:flutuacoes}(b) we consider $kTe$ = 16.2 eV, the temperature obtained through the $I \times V$ electric probe curve. Moreover, the turbulent floating potential (red cross) and the ion saturation current (blue dot) fluctuations depend on the radial position as shown by the radial profile of their turbulence levels shown in Fig. \ref{fig:flutuacoes}(c).  For each radial position presented in Fig. \ref{fig:flutuacoes}(c) we consider the local average temperature and calculate the fluctuation standard deviations ($\sigma_n$  and  $\sigma _{\varphi}$) and the mean density ($\bar{n}$) during the whole 10 s of stationary equilibrium. To obtain $\sigma _{\varphi}$ we consider the plasma potential and the floating potential fluctuations similar. Within these approximations, the two curves in Fig. \ref{fig:flutuacoes}(c) are close enough to consider the plasma adiabaticity satisfied, i. e., $\frac{\sigma_n}{\bar{n}}\cong\frac{\sigma _{\varphi}}{kTe}$.

\section{Turbulence control}
%\label{}

The turbulence level and turbulence driven particle transport are controlled by changing the radial profile of the electric field through the imposing of an external electric potential on some of the 16 available bias plates (see Fig. \ref{fig:helimak} (a)). For the shots considered in this paper, bias is imposed in four bias plates (two on the top and two on the bottom, on both sides of the machine) placed in the interval from R=0.86m to R=1.07m, as marked in yellow in Fig. \ref{fig:helimak} (a), near the radial region chosen to analyses the transport (1.1m$<$R$<$1.25m). 

\begin{figure*}
	\centering
				\includegraphics{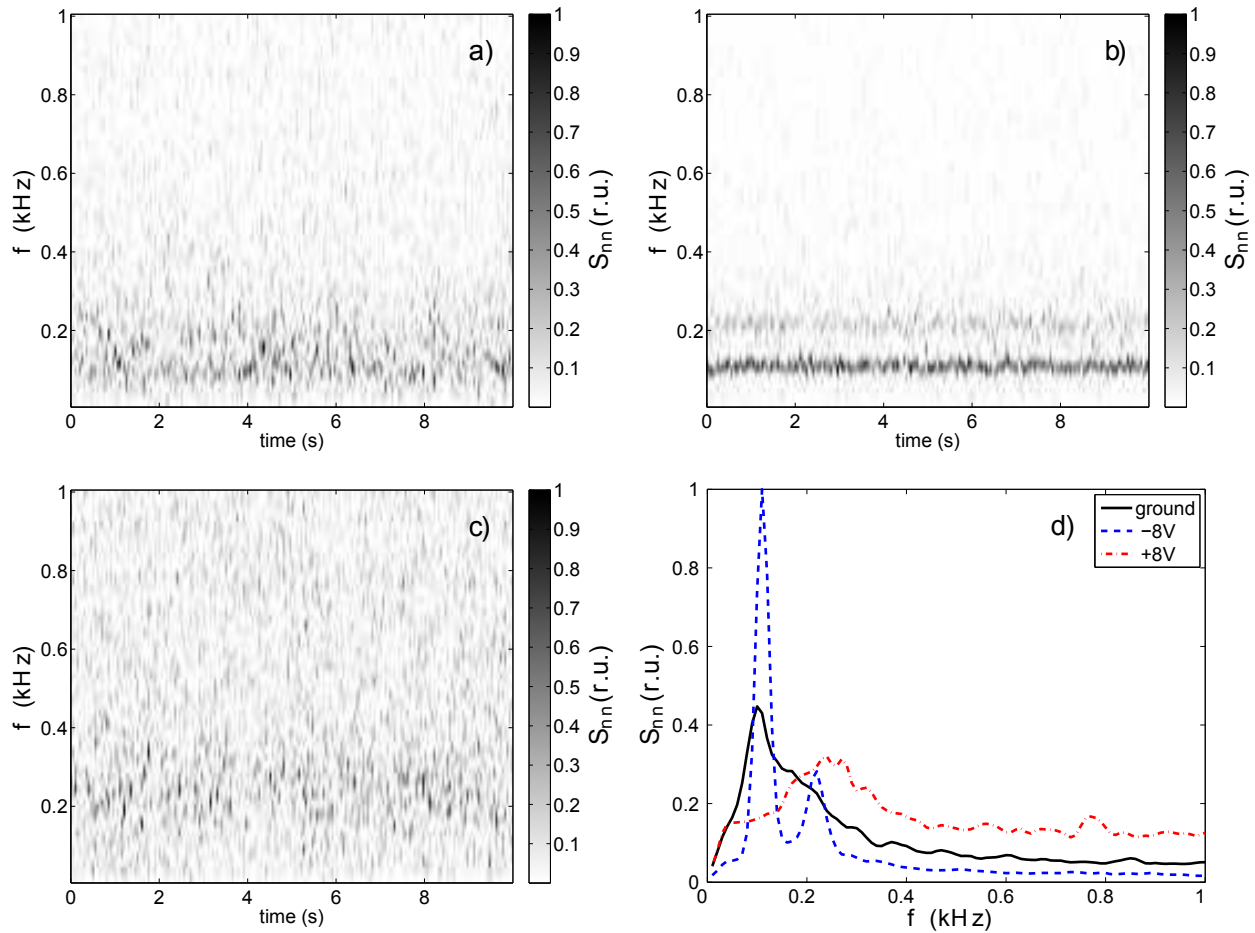}
	\caption{Evolution of saturation current power spectra for three different bias values: (a) ground, (b) -8V and (c) +8V. The three mean spectra presented in (d) are taken at the same radial position, R = 1.18m.}
	\label{fig:espectros}
\end{figure*}

Figure \ref{fig:espectros} shows an example of how the external bias changes the plasma turbulence. This figure presents the frequency spectrogram of the saturation current fluctuation for three different bias values. The spectral power, showed in grey scale, is calculated by using a windowed FFT algorithm and separately normalized for each case. The ten seconds of each time series is divided into 200 time windows with fifty thousand points each. So, for every time window the power spectrum is plotted vertically on gray scale. 

For all cases, the equilibrium plasma does not change during the whole ten seconds of the analyzed time. The spectrum for grounded bias, Fig. \ref{fig:espectros} (a), presents a concentration of power near 100 Hz with a large width.  Comparing Fig. \ref{fig:espectros} (b) obtained for bias = -8V with Fig. \ref{fig:espectros} (a), we recognize that the negative bias induces a dominant mode at frequency 100 Hz, making the power spectrum thinner. On the other hand, in Fig. \ref{fig:espectros} (c) for +8V of bias, the spectrum is broader when compared with the unperturbed spectrum.  The effect of external bias on the power spectra is better observed in Fig. \ref{fig:espectros} (d) which shows the time average of the three windowed power spectra of Fig. \ref{fig:espectros} (a-c). 

\begin{figure}
	\centering
		\includegraphics{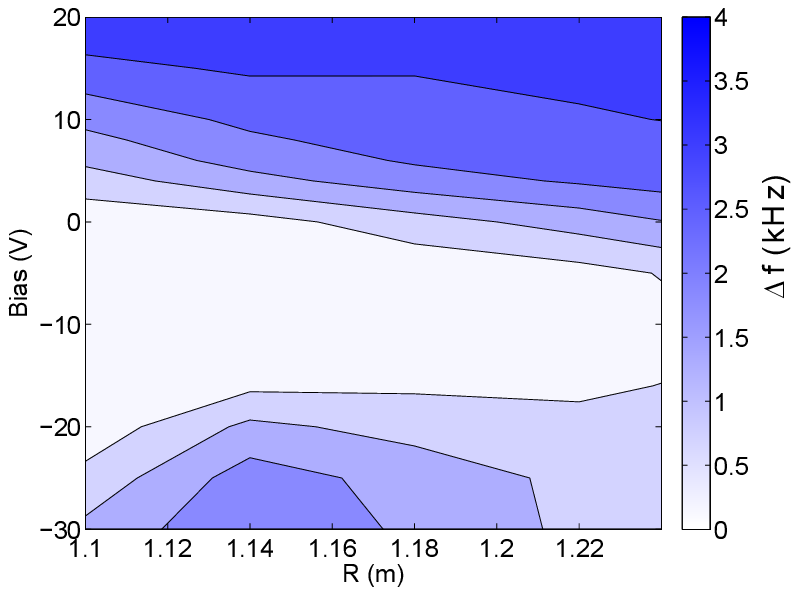}
	\caption{Radial profiles of spectral width of fluctuating saturation current for several positive and negative bias values.}
	\label{fig:deltaF}
\end{figure}

The analysis of the saturation current turbulence changes induced by the bias, presented in Fig. \ref{fig:espectros}, is extended in Fig.    \ref{fig:deltaF}, where the radial profiles of spectrum width ($\Delta f$) are calculated for several bias values. The $\Delta f$ is estimated by sorting the power spectrum channels in crescent order and then computing the number of channels that needs to be summed in order to reach half of the total spectral power.  As shown in Fig. \ref{fig:deltaF}, the external bias induces different behavior at the plasma turbulence, enlarging the broad band for positive bias and creating frequency localized modes for negative bias. Because of these different results, turbulence will be separately analyzed for positive and negative perturbing bias. 

To end the section, we point out that the adiabatic condition, shown in Figure \ref{fig:flutuacoes}(c) for the case without external biasing, is still a reasonable approximation for both negative and positive imposed bias values. %\textcolor{red}{condition l}

\section{Resonant driven transport}
\label{sec:04}

In this section we use data collected by the probe distribution shown in Fig. \ref{fig:helimak} (b) to analyze turbulence and transport alterations induced by negative bias.

We consider the spectral analysis introduced in [\onlinecite{Ritz88}] to calculate the particle transport induced by the electrostatic turbulence. Within this procedure the time average transport is given by $\Gamma = <\tilde{n} \cdot \tilde{V}_{E}>$, where $\tilde{n}$, and $\tilde{V}_{E}$ are, respectively,  the density and the $E \times B$ electric drift velocity fluctuations. As indicated in [\onlinecite{Ritz88}] we take $\tilde{n} \propto \tilde{I} _{sat}$   and $\tilde{V} _{E} \propto \tilde{E} _z$, , where $\tilde{I} _{sat}$ and $\tilde{E}_{z}$ are, respectively, the fluctuations of the ion saturation current and of the z component of the electric field.

\begin{figure}
	\centering
				\includegraphics{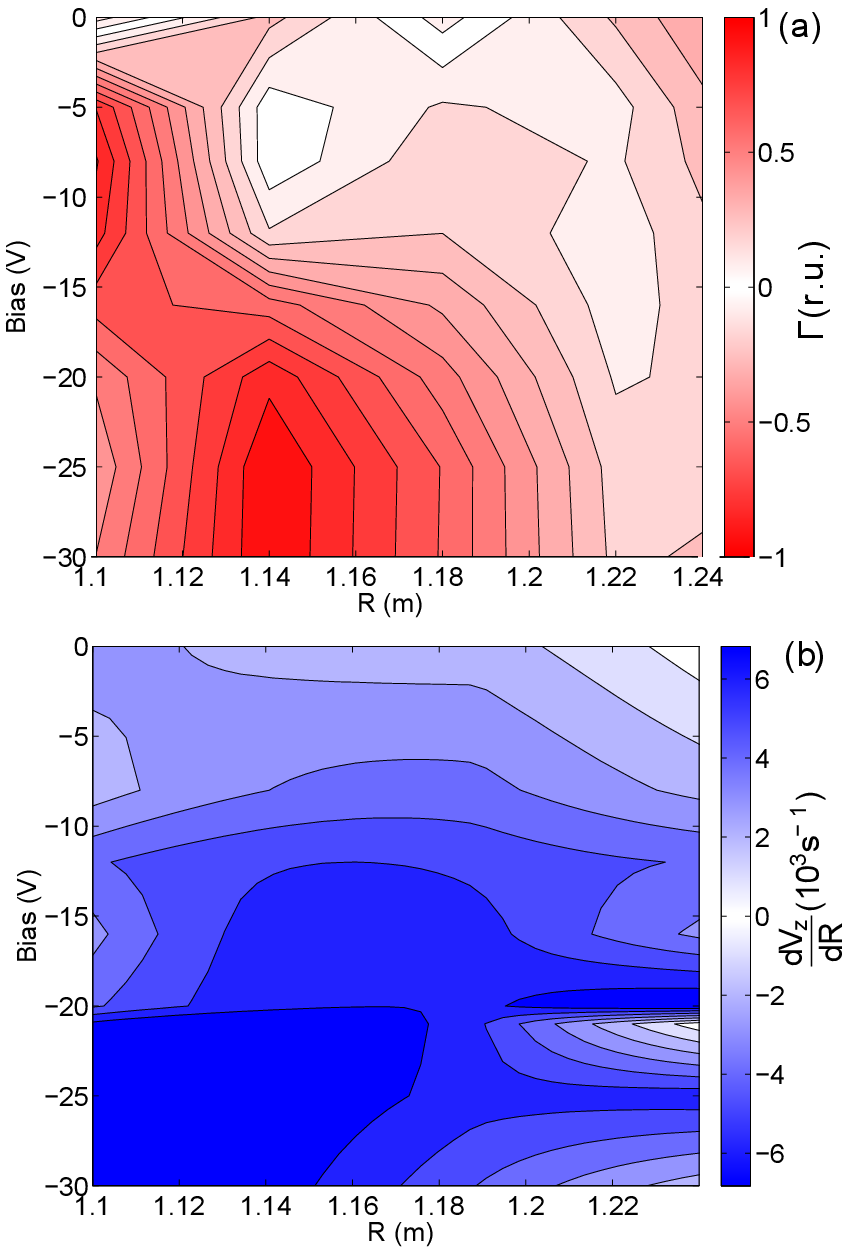}
					\caption{Radial profiles of (a) velocity shear and (b) turbulence induced radial particle transport for several negative bias values.}
	\label{fig:transp_neg}
\end{figure}

Applying the introduced spectral analysis, we calculate the radial transport profiles for several negative bias values shown in Fig. \ref{fig:transp_neg} (a). In this figure we use the same color for positive and negative transport once we are interested on the absolute values of the transport.  This figure shows that the transport depends on the radial position and the bias values. Namely, for external bias between -5V and -10V transport is high in the region around R = 1.10m, while for bias from -20V to -30V the maximum is around R = 1.14m.

In the analyzed experiments, the bias changes not only the estimated particle transport but also the plasma flow shear obtained from the spectroscopy radial flow profile. The influence of the sheared flow on the transport is predicted in [\onlinecite{Biglari90},\onlinecite{Terry00}]. A high value of velocity shear is considered useful to breaks the macroscopic structures in the plasma edge and, consequently, to reduce the transport. To check this prediction, we present in Fig. \ref{fig:transp_neg} (b) the radial profile of the velocity shear $dV_z/dR$ calculated for several external bias values. Comparing Figs. \ref{fig:transp_neg} (a) and (b), we observe, contrary to the prediction, that the transport is high in regions with high shear. Thus, the lack of transport reduction may indicates that in the analyzed region of the Texas Helimak the increase of the shear velocity does not reduce the turbulent transport.

Another effect that may contribute to the radial particle transport is the wave-particle resonant interaction, as considered in [\onlinecite{Horton85,Marcus08,Marcus08-2}]. Applying this framework to Texas Helimak, we introduce a Hamiltonian model and vertical $y$ and radial $x$ normalized coordinates. In this model, the toroidal magnetic field is uniform and the equilibrium electric field depends on the radial coordinate $x$, and two coupled drift waves propagate in the vertical direction $y$.  Thus, the plasma flow, in vertical direction, is due to the $E \times B$ equilibrium drift and the chaotic transport, in radial direction, is caused by the particle guiding center $\tilde{E} \times B$ drift. Consequently, the wave driven transport depends on the sheared plasma flow determined essentially by the radial electric field profile.

To apply the considered model we introduce an almost integrable Hamiltonian, with two small wave amplitudes such that the system is integrable if only one wave is considered:

\begin{eqnarray}
 H(x,y,t) =&&\Phi _0 (x)-V_{ph1}x+A_1sen(k_{x1}y)cos(k_{y1}y)
 \nonumber \\
 &&+A_2sen(k_{x2}x)cos\{k_{y2}[y-(V_{ph2}-V_{ph1})]\}
 \label{eq:01}
\end{eqnarray}

One of the results presented in [\onlinecite{Horton85}] is that the transport is essentially determined by the trapping profile $U$ defined as, 	
 					
 \begin{equation}
 U(x) = \frac{B_0}{A_1k_{x1}}[V_E(x)-V_{ph1}]
 \label{eq:02}
\end{equation}

In these equations $A$ is the amplitude of each wave,$V_{ph}$ is the phase velocity, $k_x$ is the radial wave number, $k_y$ is the vertical wave number, $V_E$ is the $E \times B$ drift wave velocity and $B_0$ is the uniform magnetic field.

The transport dependence on the trapping profile ($U$) can be understood by analyzing Poincaré sections of the particle guiding center drift obtained from the considered Hamiltonian.

\begin{figure}
	\centering
				\includegraphics{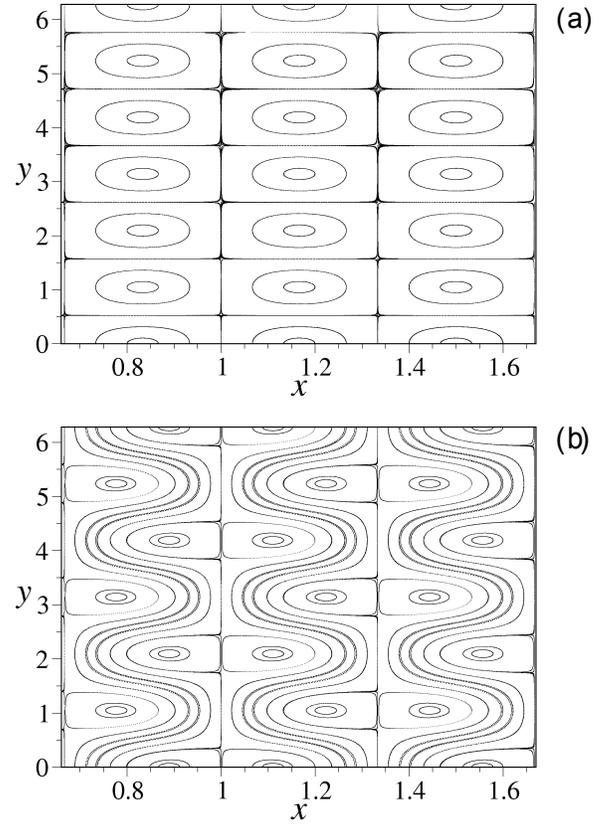}
	\caption{Phase space of the one wave integrable system for  U=0 (a) and for U=0.5 (b). On this phase space x and y are, respectively, the normalized radial and vertical coordinates.}
	\label{fig:teor_u_cte}
\end{figure}

In the observed plasma edge turbulence, the $U$ values are commonly between 0 and 1. Initially, in Fig. \ref{fig:teor_u_cte} we present phase spaces of the integrable system for one dominant wave, assuming that U is locally uniform.  For the case where U is close to zero, Fig. \ref{fig:teor_u_cte} (a), there is the formation of periodic islands separated by a separatrix in form of grid. For this equilibrium space phase, the perturbing second wave breaks the separatrix, creating chaos and causing a high transport channel along the broken separatrix. On the other hand, for a higher trapping profile, 0 $< U <$ 1, as in Fig. \ref{fig:teor_u_cte} (b), the phase space contains resonant islands separated by open lines that act as barriers, i. e.,  impede orbit displacement in the radial direction. In this last case, the perturbing second wave breaks the island separatrices and the barriers, creating chaos around the islands. However, the chaotic transport is much smaller than the one calculated for $U$ near zero.  Therefore, we can use the predicted transport dependence on the estimated equilibrium values of $U$ to explain the transport changes with bias.

\begin{figure}
	\centering
				\includegraphics{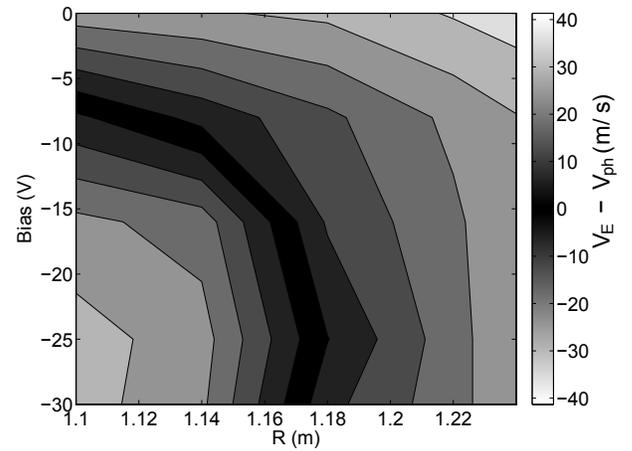}
	\caption{Radial profile of the difference between $ExB$ drift velocity and phase velocity for several negative bias values. This velocity difference is proportional to the trapping profile ($U$).}
	\label{fig:perfis_u}
\end{figure}

To interpret the observed transport profile alterations with bias, showed in Fig. \ref{fig:transp_neg} (a), we present in Fig. \ref{fig:perfis_u} the radial profile of the velocity difference $V_e-V_{ph}$ , which is directly proportional to the trapping profile $U(x)$. To calculate this velocity difference, we take the electric drift velocity $V_E$ as equal to the measured flow velocity and estimate the principal wave phase velocity $V_{ph}$ by a weighted average phase velocity of the $S(k,f)$ floating potential spectrum. The comparison between Fig. \ref{fig:perfis_u} and Fig. \ref{fig:transp_neg} (a) indicate that the transport qualitatively changes with $U(x)$ as the bias is varied. Namely, in the region where $U(x)$ is close to zero (black regions in Fig. 8) the transport is high (red regions in Fig. \ref{fig:transp_neg} (a)). Thus, our estimations indicate that the transport is mainly driven by the wave particle resonance predicted in the model considered in this section.

\section{Evidence of a shearless transport barrier}
%\label{}

\begin{figure}
	\centering
				\includegraphics{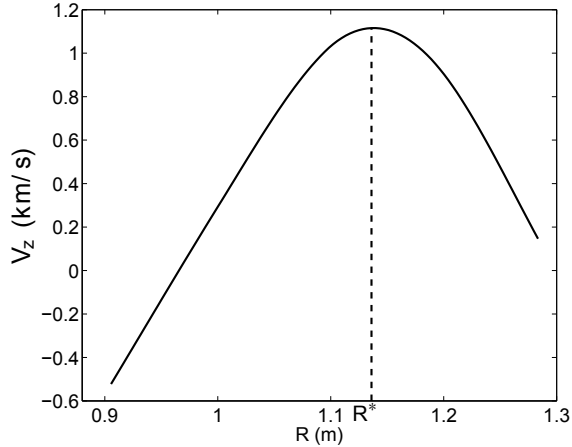}
	\caption{ Radial profile of the vertical velocity showing the radial position where the shear is zero for bias = +10V}
	\label{fig:perfil_vz}
\end{figure}

In contrast to shots analyzed in the Section \ref{sec:04}, shots perturbed by positive bias, considered in this section, present broad frequency power spectra without a dominant mode. Besides that, for discharges with a positive bias, the flow velocity radial profile presents a maximum, at R = R*, in the analyzed radial interval. Thus, for +10V of external bias, Fig. \ref{fig:perfil_vz} shows an example of velocity radial profile with a maximum at R* = 1.13m. The presence of this maximum implies that the plasma has a shearless region at R* and reverses its shear flow around this radial position. 

Transport studies in reversed shear flows in fluids [\onlinecite{Sommeria89},\onlinecite{Solomon93}] and plasmas [\onlinecite{Marcus08-2},\onlinecite{Negrete00}] reveal the presence of a transport barrier in the shearless radial position that impede or reduce the radial transport in the reversed shear region. In our work, to examine the effect of the reversed shear on the transport during discharges perturbed by positive bias, we perform a directly comparison between the velocity shear and the electrostatic turbulence induced transport. 

\begin{figure}
	\centering
				\includegraphics{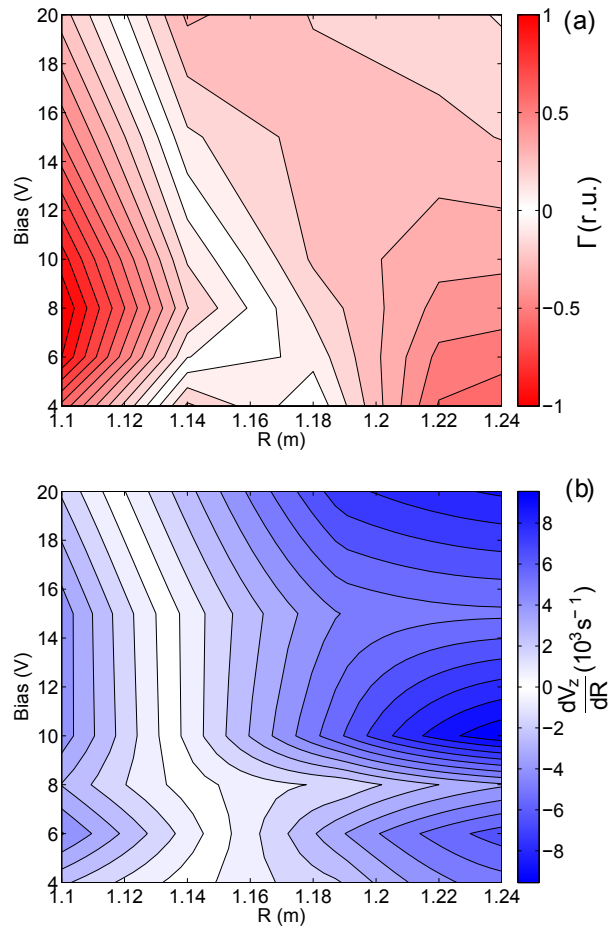}
					\caption{Radial profiles of (a) velocity shear and (b) turbulence induced radial particle transport for several positive bias values.}
	\label{fig:transp_pos}
\end{figure}

To verify how the transport depends on the radial position and the applied positive bias values, we perform the same spectral analysis introduced in section 4. Thus, we show in Fig. \ref{fig:transp_pos} (a) the transport radial profile for several positive bias values. In this figure we use the same color for positive and negative transport. Analogous to Fig. \ref{fig:transp_neg} (b), we show in Fig. \ref{fig:transp_pos} (b) the radial profile of the velocity shear $dV_z/dR$ calculated for several external positive bias values. For all analyzed bias values, Fig. \ref{fig:transp_pos} shows that the particle transport is very small where the velocity shear is null. This observation can be interpreted as an indication of a shearless barrier.

\begin{figure}
	\centering
				\includegraphics{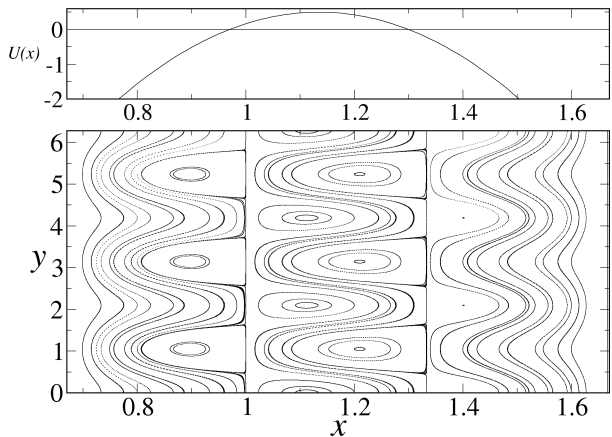}
	\caption{Phase space of the integrable system, for a wave with the average phase velocity,  obtained for a non monotonic trapping profile U(x). On this phase space x and y are, respectively, the normalized radial and vertical coordinates}
	\label{fig:teor_u_var}
\end{figure}

We can adapt the Hamiltonian model introduced in Section \ref{sec:04} to interpret the observed transport reduction in the shearless region. Thus, to obtain the Poincaré section of the guiding center orbits in phase space, necessary to our interpretation, we consider the nonmonotonic trapping profile $U(x)$ determined by the flow profile and the average phase velocity. In Fig. \ref{fig:teor_u_var}, for a bias value of +10V, we present one example of such Poincaré sections obtained for a $U(x)$ shape (shown in Fig. \ref{fig:teor_u_var}) estimated from the flow velocity shown in Fig. \ref{fig:perfil_vz}.

The results illustrated in Fig. \ref{fig:teor_u_var} give an indication of a transport barrier located near the shearless position, at $x \approx$ 1.2, where $U(x)$ presents a maximum. Consequently, the barrier identified in Fig. \ref{fig:teor_u_var} can explain the transport reduction in the sherless region (see Fig. \ref{fig:transp_pos} (a)). Moreover, one must note that, for all positive bias, the two white regions in Figs. \ref{fig:transp_pos} (a) and (b) clearly indicate a transport reduction in the shearless region.

\section{Conclusions}
%\label{}

In this work we analyzed the particle transport in Texas Helimak for external biased discharges with fluctuations obtained from a large set of Langmuir probes and a sheared flow velocity  measured by a spectroscopy diagnostic. We showed that the external bias value changes spectral plasma characteristics, inducing a frequency localized mode in negative bias values and a broad band spectrum in positive bias values. 

We calculated the turbulence induced particle transport for several negative bias values and the obtained results show that the transport in Texas Helimak is not reduced in the high velocity shear regions. On the other hand, for these discharges, the transport profile has a maximum localized near the region where the $E \times B$ drift velocity is approximately equal to the wave phase velocity. A particle-wave resonant interaction model is used to interpret these maxima transport regions as a resonant effect that increases the transport.

For positive bias values the flow velocity radial profiles have a maximum inside the analyzed radial region, so the velocity shear is zero where the velocity reaches a maximum and reverses its signal near this position. We observe that the transport profiles are close to zero in the same radial positions where the velocity shear is zero. This observation is interpreted as evidence of a shearless transport barrier expected in reversed shear flow systems.

In conclusion, our analysis of the transport in Texas Helimak reveal new features that may contribute to improve the understanding of the particle transport and their reduction associated to the barrier onset in plasma edge fusion machines.

%\begin{acknowledgments}
% Put your acknowledgments here.
%\end{acknowledgments}
%\bibliography{bib_pop_transp}{}


\end{document}